\documentclass[twocolumn,showpacs,preprintnumbers,pra,hyperref]{revtex4}%
\usepackage{amsfonts}
\usepackage{amsmath}
\usepackage{amssymb}
\usepackage{graphicx}%
\begin{document}
\title{Goos-H\"{a}nchen-Like Shifts in Atom Optics}
\author{Jianhua Huang,$^{1}$ Zhenglu Duan,$^{1}$ Hong Y. Ling,$^{2,\dag}$ and Weiping
Zhang$^{1,\ast}$} \affiliation{$^{1}$State Key Laboratory of
Precision Spectroscopy,} \affiliation{Department of Physics, East
China Normal University, Shanghai 200062, P. R. China,}
\affiliation{$^{2}$Department of Physics and Astronomy, Rowan
University, Glassboro, New Jersey 08028-1700}

\begin{abstract}
We consider the propagation of a matter wavepacket of two-level atoms through
a square potential created by a super-Gaussian laser beam. \ We explore the
matter wave analog of Goos-H\"{a}nchen shift within the framework of atom
optics where the roles of atom and light is exchanged with respect to
conventional optics. \ Using a vector theory, where atoms are treated as
particles possessing two internal spin components, we show that\ not only
large negative but also large positive Goos-H\"{a}nchen shifts can occur in
the reflected atomic beam.

\end{abstract}

\maketitle

\section{Introduction}

In conventional optics for light waves, Goos-H\"{a}nchen in 1947
discovered that a light beam under the condition of total reflection
can experience a lateral shift (or displacement) along the surface
of a dielectric boundary \cite{Goos}. \ \  This pioneering work has
stimulated a large volume of studies
\cite{Artmann,Chiao,Li,WangLG:3,Xiang,Zeng,Renard,
Tamir,Wild,Schlesser,Pfleghaar,Lai:2,Bonnet,Lai:1,Berman,Broe,Kong,Fan,Declercq,
Lakhtakia,Shadrivov,Felbacq,Qing,Chen,WangLG:1,WangLG:2,Shen,Liu:2,He,Tsakmakidis,Yan}%
 \ , concerning the
Goos-H\"{a}nchen shift of reflected or transmitted
\cite{Chiao,Li,WangLG:3} light beams of different polarizations
\cite{Lai:2} in different media characterized with, for example,
periodic structures \cite{Bonnet,Declercq,Felbacq,He}, (left or
right) handedness \cite{Shadrivov,Qing,Chen,WangLG:2}, multilayers
\cite{Tamir}, weakly absorbing \cite{Lai:1,Shen}, lower-index
\cite{Fan} or negative-index of refraction \cite{Berman}, etc. \ The
key physics behind the Goos-H\"{a}nchen shift is the nature of wave
interference. \ From the perspective of wave optics, the incident
beam of a finite transverse width can be viewed as composed of plane
wave components, each of which has a slightly different transverse
wavevector. \ Each wave component, after the total internal
reflection, undergoes a different phase shift, and the superposition
of all the reflected wave components gives rise to the lateral shift
of the intensity peak in the reflected beam \cite{Artmann}.

In this sense, it is not so much the total internal reflection but
rather the phase modulations for different plane wave components
that remains the true mechanism behind the lateral shift. \ Thus,
the Goos-H\"{a}nchen shift is expected to occur in matter waves
where particles have finite masses. \ As is known, under the usual
conditions (or temperatures), electrons possess a de Broglie's
wavelength much longer than atoms because the latter is much heavier
in mass than the former.\ \ Thus, it is much easier to demonstrate
the Goos-H\"{a}nchen shift with electrons \cite{Miller,Fradkin} or
even neutrons \cite{Maaza,Ignatovich} than with atoms. The
situation, however, has been rapidly changed over the last two
decades. Nowadays, ultracold atoms with a relatively long de
Broglie's wavelength can be routinely made available, thanks to the
rapid advancement of the laser cooling and trapping technology. \
Motivated by the fact that ultracold atoms have led to many
important applications in atom optics \cite{Meystre}, we explore, in
this paper, the matter wave analog of Goos-H\"{a}nchen effect within
the framework of atom optics where matter waves of ultracold atoms
are manipulated by laser fields. \ \ An important difference between
the matter waves in atom optics and the light waves in conventional
optics is that atoms have internal electronic structures while
photons are structureless. \ Thus, an accurate description of the
Goos-H\"{a}nchen effect in atom optics must regard atoms as
particles possessing internal spins (energy states). \ To the best
of our knowledge, our work here represents the first that is
seriously devoted to the problem of Goos-H\"{a}nchen effect with
cold atoms. \ As such, we limit our goals to establishing a general
theoretical framework and to applying it for a basic understanding
of the Goos-H\"{a}nchen effect in matter waves with cold two-level
atoms, while at the same time hoping that our work can draw
significant attentions from experimentalists for future
applications.

Our paper is organized as follows. \ In Sec. \ref{model}, we derive
a set of coupled 1-D Schr\"{o}dinger equations to describe the
scattering of two-level atoms by a super-Gaussian laser beam in a
2-D setting. \ In the same section, we\ present the connection
between the lateral shifts and the coefficients of reflection and
transmission. \ In Sec. \ \ref{vector case}, we derive, with the
help of a dressed state picture, a set of analytical expressions for
the reflection and transmission coefficients, which are to be used\
in Sec. \ref{discussion} to significantly simplify our calculations.
\ In Sec. \ref{discussion}, we combine the tools developed in Sec.
\ref{model} with those in Sec. \ref{vector case} to numerically
investigate, within the context of atom optics, the matter wave
analog of Goos-H\"{a}nchen-like shifts. \ Finally, a conclusion will
be given in Sec. \ref{conclusion}.

\section{Model and Basic Equations}

\label{model}

\begin{figure}[h]
\centering
\resizebox{7cm}{!}{\includegraphics{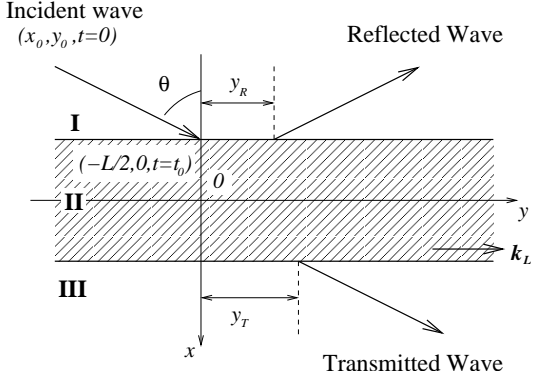}}\caption{A wave
packet of atoms with two internal states impinges on a laser
\textquotedblleft
slab\textquotedblright.}%
\label{fig1}%
\end{figure}

Figure \ref{fig1} is the schematic of our model, in which a matter
wavepacket composed of two-level atoms of transition frequency
$\omega_{a}$ is obliquely incident upon a \textquotedblleft
slab\textquotedblright\ \ made up of a travelling laser beam of
frequency $\omega_{L}$ and wavenumber $k_{L}$.\ In our model, we
require that both the atom and laser beams be sufficiently wide
along the direction normal to the plane of incidence ($x-y$ plane)
so that the degree of freedom in the $z$-dimension can be completely
liberated. \ Under such a circumstance, we can adopt the following
coupled $2-D$ Schr\"{o}dinger equations \cite{Zhang}
\begin{subequations}
\label{SE}%
\begin{align}
i\dot{\Psi}_{1}  &  =-\frac{\hbar\nabla^{2}}{2m}\Psi_{1}-\frac{\Omega\left(
x\right)  }{2}e^{-ik_{L}y}\Psi_{2},\\
i\dot{\Psi}_{2}  &  =\left(  -\frac{\hbar\nabla^{2}}{2m}-\Delta-i\frac{\gamma
}{2}\right)  \Psi_{2}-\frac{\Omega\left(  x\right)  }{2}e^{ik_{L}y}\Psi_{1},
\end{align}
\end{subequations}
to describe the evolution of the wavefunctions, $\Psi_{1}$ and
$\Psi_{2}$, of the ground state $\left\vert 1\right\rangle $ and the
excited state $\left\vert 2\right\rangle $. \ In Eqs. (\ref{SE}), we
have defined $m$ as the atomic mass,
$\nabla^{2}\equiv\partial^{2}/\partial x^{2}+\partial
^{2}/\partial y^{2}$ as the 2-D Laplacian operator, $\Delta=\omega_{L}%
-\omega_{a}$ as the laser detuning, and $\gamma$ as the decay rate
of the excited atomic state. \ In addition, we describe the dipole
interaction between the laser field and atoms by a Rabi frequency in
the form of
\begin{equation}
\Omega\left(  x\right)  =\Omega F\left(  x\right)  , \label{Omega}%
\end{equation}
where $\Omega$ is the peak value and $F(x)$ is a normalized spatial function,
representing the laser profile. \ In this paper, the laser is assumed to
possess a beam profile in the form of a high-order Gaussian function,
$F\left(  x\right)  =\exp\left[  -\left(  x^{2}/w_{L}^{2}\right)  ^{N}\right]
$; such a beam is experimentally accessible through spatial shaping techniques
\cite{Liu:1,Belanger,Dong} or optical techniques \cite{Zhao}. \ To further
simplify the problem, we restrict our study to the super-Gaussian beam with an
order number $N$ so large that, to a fairly good approximation, $F\left(
x\right)  $ can be idealized as a step function%
\begin{equation}
F\left(  x\right)  =%
\begin{cases}
1, & -L/2<x<L/2,\\
0, & x>L/2,x<-L/2,
\end{cases}
\label{Omega style 2}%
\end{equation}

Next, we utilize the fact of the Rabi frequency [Eq. (\ref{Omega})] being $y$
independent to eliminate $y$ in favor of $y$-wavevector $k_{y}$ through the
Fourier transformation
\begin{subequations}
\label{state}%
\begin{align}
\Psi_{1}\left(  \mathbf{r},t\right)   &  =\int dk_{y}\phi_{1}\left(
x,k_{y},t\right)  e^{ik_{y}y-\frac{i\hbar k_{y}^{2}}{2m}t},\\
\Psi_{2}\left(  \mathbf{r},t\right)   &  =\int dk_{y}\phi_{2}\left(
x,k_{y},t\right)  e^{i(k_{y}+k_{L})y-\frac{i\hbar k_{y}^{2}}{2m}t},
\end{align}
\end{subequations}
where $\mathbf{r=}\left(  x,y\right)  $. \ By doing so, we transform
Eqs.
(\ref{SE}) into coupled 1-D Schr\"{o}dinger equations%

\begin{equation}
i\hbar\mathbf{\dot{\phi}=}\left(  -\frac{\hbar^{2}}{2m}\frac{\partial^{2}%
}{\partial x^{2}}\hat{I}+\hat{V}\right)  \mathbf{\phi,} \label{xSE}%
\end{equation}
where $\mathbf{\phi}=\left(  \phi_{1},\phi_{2}\right)  ^{T}$ is a
two-component vector field, $\hat{I}\,\ $is a $2\times2$ unit matrix, and
$\hat{V}$ is the potential matrix given by
\begin{equation}
\hat{V}=-\frac{\hbar}{2}\left[
\begin{array}
[c]{cc}%
0 & \Omega\\
\Omega & 2\left(  \delta+\frac{1}{2}i\gamma\right)
\end{array}
\right]  . \label{V_M}%
\end{equation}
In Eq. (\ref{V_M}), we have defined%
\begin{equation}
\delta=\Delta-\frac{\hbar k_{y}k_{L}}{m}-\frac{\hbar k_{L}^{2}}{2m},
\label{detuning}%
\end{equation}
as the effective detuning, where $\hbar k_{y}k_{L}/m$ and $\hbar k_{L}^{2}/2m$
are, respectively, the Doppler and the photon recoil frequency. \ Equation
(\ref{xSE}) serves as the starting point for next section, where we calculate
the scattering matrix, which determines all the scattering properties of our model.

For now, we turn our attention to the lateral shifts of the reflected and
transmitted wavepackets. \ For this purpose, \ let's first jump ahead to Eqs.
(\ref{phi1phi3}) of Sec. \ref{vector case}, which define various transmission
and reflection coefficients via the stationary scattering solutions in free
space. \ Of relevance to our interest here are the coefficients of
transmission $T_{1}$ and reflection $R_{1}$ of the ground state; we ignore
$T_{2}$ and$\ R_{2}$ of the excited state since the excited wave, being highly
susceptible to the decay by the spontaneous emission, cannot propagate far
from the scattering region. \ Let $\Theta_{1}^{R,T}\left(  \mathbf{k}\right)
$ be the phases of reflection and transmission coefficients defined through
the relation
\begin{equation}
S_{1}\left(  \mathbf{k}\right)  =\left\vert S_{1}\left(  \mathbf{k}\right)
\right\vert e^{i\Theta_{1}^{S}\left(  \mathbf{k}\right)  }, \label{theta_RT}%
\end{equation}
where for notational simplicity, we have used (and will continue to use) $S=R$
and $S=T$ to symbolize reflection and transmission, respectively. \ For an
incident wave packet initially ($t=0$) located at $\left(  x_{0}\text{, }%
y_{0}\right)  $ far away from the laser beam (see Fig. \ref{fig1}), we can
construct, through the superposition of the time-independent solutions [Eqs.
(\ref{phi1phi3})], its reflected and transmitted wavepackets in the form of%

\begin{equation}
\Psi_{1}^{S}\left(  \mathbf{r},t\right)  =\int d^{2}\mathbf{k}^{\prime
}f\left(  \mathbf{k}^{\prime}-\mathbf{k}\right)  |S_{1}\left(  \mathbf{k}%
^{\prime}\right)  |e^{i\eta_{1}^{S}\left(  \mathbf{k}^{\prime},\mathbf{r}%
\right)  } \label{psi_RT}%
\end{equation}
where $\eta_{1}^{R,T}$ are the total phases defined as
\begin{subequations}
\label{phase_RT}%
\begin{align}
\eta_{1}^{R}\left(  \mathbf{k}^{\prime},\mathbf{r}\right)   &  =\Theta_{1}%
^{R}\left(  \mathbf{k}^{\prime}\right)  -k_{x}^{\prime}\Delta x^{R}%
+k_{y}^{\prime}\Delta y^{R}-\frac{\hbar{k}^{\prime{2}}t}{2m},\\
\eta_{1}^{T}\left(  \mathbf{k}^{\prime},\mathbf{r}\right)   &  =\Theta_{1}%
^{T}\left(  \mathbf{k}^{\prime}\right)  +k_{x}^{\prime}\Delta x^{T}%
+k_{y}^{\prime}\Delta y^{T}-\frac{\hbar{k}^{\prime}{}^{2}t}{2m},
\end{align}
\end{subequations}
with$\ \Delta x^{R}=x+x_{0}$, $\Delta x^{T}=x-x_{0}$, $\Delta
y^{R}=\Delta
y^{T}=y-y_{0}$, and $k^{\prime2}={k}^{\prime}{_{x}}^{2}+{k}^{\prime}{_{y}}%
^{2}$. \ In Eqs. (\ref{psi_RT}), $f\left(  \mathbf{k}^{\prime}-\mathbf{k}%
\right)  $ is a (real) weighting function peaked around
$\mathbf{k}^{\prime }=\mathbf{k}$ with a momentum distribution
sufficiently narrow and smooth that Eqs. (\ref{psi_RT}) represent\
fairly accurately the plane waves of velocity $\hbar\mathbf{k}/m$. \
Equation (\ref{psi_RT}) indicates that the reflected and transmitted
waves are the result of interference among different wave components
distinguished by wavevector $\mathbf{k}^{\prime}=\left(
k_{x}^{\prime},k_{y}^{\prime}\right)  $. \ As a result, the values
of these waves at a given time and location, $\left(
\mathbf{r},t\right)  $, depend crucially on the phases [Eqs.
(\ref{phase_RT})] of each $\mathbf{k}^{\prime}$ component. \ In
particular, $\Psi_{1}^{R,T}\left(  \mathbf{r},t\right)  $ reach peak
values when a constructive interference takes place or equivalently
when $\nabla_{\mathbf{k}^{\prime}}\eta_{1}^{R,T}$ at
$\mathbf{k}^{\prime }=\mathbf{k}$ vanish. \ \ Using this condition,
we find that the peaks of the wavepackets in the coordinate space
propagate with time according to
\begin{subequations}
\label{Delta_xy}%
\begin{align}
\Delta x^{R}  &  =-\frac{\hbar k_{x}}{m}t+\frac{\partial\Theta_{1}^{R}%
}{\partial k_{x}},\ \Delta y^{R}=\frac{\hbar k_{y}}{m}t-\frac{\partial
\Theta_{1}^{R}}{\partial k_{y}},\\
\Delta x^{T}  &  =+\frac{\hbar k_{x}}{m}t-\frac{\partial\Theta_{1}^{T}%
}{\partial k_{x}},\ \Delta y^{T}=\frac{\hbar k_{y}}{m}t-\frac{\partial
\Theta_{1}^{T}}{\partial k_{y}},
\end{align}
\end{subequations}
where all the derivatives are assumed to be taken at
$\mathbf{k}^{\prime }=\mathbf{k}$.

Let $t_{0},$ $t_{R}$ and $t_{T}$ be, respectively, the time duration of the
atomic beam between $t=0$ and right before it hits the boundary at $x=-L/2$,
between $t=0$ and immediately after it is reflected from the boundary at
$x=-L/2$, and between $t=0$ and right after it is transmitted from the
boundary at $x=L/2$. \ In terms of $t_{0}$, we have $x_{0}=-(L/2+\hbar
k_{x}t_{0}/m)$ and $y_{0}=-\hbar k_{y}t_{0}/m$, which, when substituted into
Eqs. (\ref{Delta_xy}) for $\Delta x^{R,T}$, allows us to find $t_{R,T}$%
\begin{equation}
\Delta t_{S}=t_{S}-t_{0}=\frac{m}{\hbar k_{x}}\left(  \frac{\partial\Theta
_{1}^{S}}{\partial k_{x}}+L\right)  . \label{shiftt}%
\end{equation}
Finally, by incorporating these results into Eqs. (\ref{Delta_xy}) for $\Delta
y^{R,T}$, we find that the induced Goos-H\"{a}nchen-like lateral shift $y_{R}$
due to the reflection and $y_{T}$ due to the transmission (see Fig.
\ref{fig1}) are governed by
\begin{equation}
y_{S}=\frac{\hbar k_{y}}{m}\Delta t_{S}-\frac{\partial\Theta_{1}^{S}}{\partial
k_{y}}. \label{shifts}%
\end{equation}
where $\ \partial\Theta_{1}^{S}/\partial k_{i}$ ($i=x,y$ and $S=R,T$) are
evaluated using \
\begin{equation}
\frac{\partial\Theta_{1}^{S}}{\partial{k_{i}}}=-i\left[  \frac{1}{S_{1}}%
\frac{\partial S_{1}}{\partial{k_{i}}}-\frac{1}{\left\vert S_{1}\right\vert
}\frac{\partial\left\vert S_{1}\right\vert }{\partial{k_{i}}}\right]  ,
\label{thetac-R}%
\end{equation}
\ which is a direct consequence of Eq. (\ref{theta_RT}). \ In contrast to the
usual shifts, which are solely determined by the part directly proportional to
$\Delta t_{S}$, the lateral shifts in Eq. (\ref{shifts}) contain an additional
term $\partial\Theta_{1}^{S}$/$\partial k_{y}$. \ This is a unique aspect of
atom optics, where momentum conservation during the photon emission and
absorption makes the effective laser detuning $\delta$ [Eq. (\ref{detuning})]
$k_{y}$ dependent. \ \ As a result, the phase $\Theta_{1}^{S}$ becomes a
function of $k_{y}$ via its dependence on $\delta$, which, in turn, leads to a
finite\ $\partial\Theta_{1}^{S}$/$\partial k_{y}$. \

From this derivation, it is clear that (a) the Goos-H\"{a}nchen-like lateral
shifts are the wave phenomena, that depend crucially on the ability of the
optical potential to modify the phases of various matter wave components, and
(b) the key to the lateral shifts is the transmission and reflection
coefficients, which will be the focus of our study in the next section.

\section{Transmission and Reflection Coefficients}

\label{vector case}

In this section, we construct the reflection and transmission coefficients,
starting from the stationary scattering solutions of Eq. (\ref{xSE}) for an
incident ground atomic beam having an energy $E_{x}=\hbar k_{x}^{2}/2m$ and
wavenumber $k_{x}$ along the $x$ dimension. \ Let's first introduce the
reflection and transmission coefficients for the ground state, $R_{1}$ and
$T_{1}$, and those for the excited state, $R_{2}$ and $T_{2}$, via the
scattering solutions in regions I and III. \ By virtue of the decoupling
between the excited and ground states in free propagation regions I and III
outside the laser slab, the scattering solutions take the form%

\begin{subequations}
\label{phi1phi3}%
\begin{align}
\mathbf{\phi}^{I}  &  =\left(
\begin{array}
[c]{c}%
e^{ik_{1}x}+R_{1}e^{-ik_{1}x}\\
R_{2}e^{-ik_{2}x}%
\end{array}
\right)  e^{-i\frac{E_{x}}{\hbar}t},\\
\mathbf{\phi}^{III}  &  =\left(
\begin{array}
[c]{c}%
T_{1}e^{ik_{1}x}\\
T_{2}e^{ik_{2}x}%
\end{array}
\right)  e^{-i\frac{E_{x}}{\hbar}t},
\end{align}
\end{subequations}
where we have defined the free-space wavevectors
\begin{equation}
k_{1}=k_{x},k_{2}=\sqrt{2\frac{m}{\hbar}\left(  \delta+i\frac{\gamma}%
{2}\right)  +k_{x}^{2}}. \label{k1k2}%
\end{equation}

The excited-state and ground-state components in region II are, however, mixed
because Eq. (\ref{xSE}) is a coupled equation. To solve Eq. (\ref{xSE}) and
thus, to find the vector wavefunction $\mathbf{\phi}^{II}$ in region II, we
first seek to diagonalize the matrix $\hat{V}$ [Eq. (\ref{V_M})] by looking
for the eigenvectors of $\hat{V}$. \ \ This leads to two eigenvalues $V_{+}$
and $V_{-}$, given by%
\begin{equation}
V_{\pm}=\frac{\hbar}{2}\left[  -\left(  \delta+i\frac{\gamma}{2}\right)
\pm\sqrt{\left(  \delta+i\frac{\gamma}{2}\right)  ^{2}+\Omega^{2}}\right]  .
\label{Vn}%
\end{equation}
The corresponding eigenvectors are expressed as the first and second column
vectors of the following transformation matrix
\begin{equation}
U=\left(
\begin{array}
[c]{cc}%
\sin\varphi & \cos\varphi\\
-e^{i\beta}\cos\varphi & e^{-i\beta}\sin\varphi
\end{array}
\right)  , \label{U}%
\end{equation}
where $\varphi$ and $\beta$, defined as
\begin{equation}
\tan\varphi=\frac{\hbar\Omega}{2\left\vert V_{+}\right\vert },V_{+}=\left\vert
V_{+}\right\vert e^{i\beta}, \label{alpha}%
\end{equation}
are two angles introduced to characterize the dressed states%

\begin{subequations}
\label{dressedst}%
\begin{align}
\left\vert +\right\rangle  &  =\sin\varphi\left\vert 1\right\rangle
-e^{i\beta}\cos\varphi\left\vert 2\right\rangle ,\\
\left\vert -\right\rangle  &  =\cos\varphi\left\vert 1\right\rangle
+e^{-i\beta}\sin\varphi\left\vert 2\right\rangle ,
\end{align}
\end{subequations}

The inverse of $U$, which will also be an important part of the
scattering problem involving vector matter waves, is given by
\begin{equation}
U^{-1}=f\left(
\begin{array}
[c]{cc}%
e^{-i\beta}\sin\varphi & -\cos\varphi\\
e^{i\beta}\cos\varphi & \sin\varphi
\end{array}
\right)  \label{U-}%
\end{equation}
where $f=\left(  e^{-i\beta}\sin^{2}\varphi+e^{i\beta}\cos^{2}\varphi\right)
^{-1}$ is a normalization factor. \ In the absence of spontaneous decay
$\gamma=0$, we have $\beta=0$, and $U$ becomes unitary. The dressed states can
be simplified into the ones well-known in quantum optics \cite{Jaynes}. \ The
presence of the spontaneous decay renders the Hamiltonian nonhermitian
\cite{Delgado}, which is why $U$ is no longer a unitary matrix when $\beta$
$\neq0$ as one can easily verify from Eq. (\ref{U}).

With these preparations, we now express $\mathbf{\phi}^{II}$, in terms of
wavefunctions $\phi_{\pm}$ on the dressed-state basis, as%
\begin{align}
\mathbf{\phi}^{II}  &  =U\left(
\begin{array}
[c]{c}%
\phi_{+}\\
\phi_{-}%
\end{array}
\right)  e^{-i\frac{E_{x}}{\hbar}t},\nonumber\\
&  \equiv U\left(
\begin{array}
[c]{c}%
A_{1}e^{\alpha_{1}x}+B_{1}e^{-\alpha_{1}x}\\
A_{2}e^{\alpha_{2}x}+B_{2}e^{-\alpha_{2}x}%
\end{array}
\right)  e^{-i\frac{E_{x}}{\hbar}t}, \label{phi2}%
\end{align}
where $A_{i}$ and $B_{i}$ are\ the superposition coefficients, and for easy
organization, we introduce $\alpha_{1}\equiv\alpha_{+}$ and $\alpha_{2}%
\equiv\alpha_{-}$, where
\begin{equation}
\alpha_{\pm}=\sqrt{2mV_{\pm}/\hbar^{2}-k_{x}^{2}}. \label{alpha+-}%
\end{equation}
\ To facilitate the derivation bellow, besides $U$ \ and $U^{-1}$, we also
introduce the matrix $W$ and its inverse $W^{-1},$ where%
\begin{equation}
W=\left(
\begin{array}
[c]{cc}%
ik_{1} & 0\\
0 & ik_{2}%
\end{array}
\right)  ^{-1}U,W^{-1}=U^{-1}\left(
\begin{array}
[c]{cc}%
ik_{1} & 0\\
0 & ik_{2}%
\end{array}
\right)  \label{WW-}%
\end{equation}
or equivalently,
\begin{equation}
W_{ij}=\left(  ik_{i}\right)  ^{-1}U_{ij},\left(  W^{-1}\right)  _{ij}%
=ik_{j}\left(  U^{-1}\right)  _{ij}\text{. \ } \label{Wij}%
\end{equation}
Next, we require that $\phi^{I,II,III}$ and their derivatives be continuous at
location $x=L/2,$ leading to%

\begin{subequations}
\label{uv}%
\begin{align}
\left(
\begin{array}
[c]{c}%
u_{1}\\
u_{2}%
\end{array}
\right)   &  =U^{-1}\left(
\begin{array}
[c]{c}%
T_{1}e^{ik_{1}L/2}\\
T_{2}e^{ik_{2}L/2}%
\end{array}
\right)  ,\\
\left(
\begin{array}
[c]{c}%
v_{1}\\
v_{2}%
\end{array}
\right)   &  =W^{-1}\left(
\begin{array}
[c]{c}%
T_{1}e^{ik_{1}L/2}\\
T_{2}e^{ik_{2}L/2}%
\end{array}
\right)
\end{align}
\end{subequations}
where $\left(  u_{i},v_{i}\right)  $ are a set of new variables
defined as
\begin{subequations}
\label{ui vi}%
\begin{align}
u_{i}  &  =A_{i}e^{\alpha_{i}\frac{L}{2}}+B_{i}e^{-\alpha_{i}\frac{L}{2}},\\
v_{i}  &  =\alpha_{i}\left(  A_{i}e^{\alpha_{i}\frac{L}{2}}-B_{i}%
e^{-\alpha_{i}\frac{L}{2}}\right)
\end{align}
\end{subequations}
Similarly, application of the continuation conditions at location
$x=-L/2$ results in
\begin{subequations}
\label{XY1}%
\begin{align}
U\left(
\begin{array}
[c]{c}%
x_{1}\\
x_{2}%
\end{array}
\right)   &  =\left(
\begin{array}
[c]{c}%
e^{-ik_{1}L/2}+R_{1}e^{ik_{1}L/2}\\
R_{2}e^{ik_{2}L/2}%
\end{array}
\right)  ,\label{X}\\
W\left(
\begin{array}
[c]{c}%
y_{1}\\
y_{2}%
\end{array}
\right)   &  =\left(
\begin{array}
[c]{c}%
e^{-ik_{1}L/2}-R_{1}e^{ik_{1}L/2}\\
-R_{2}e^{ik_{2}L/2}%
\end{array}
\right)  , \label{Y}%
\end{align}
\end{subequations}
where $\left(  x_{i}\text{,}y_{i}\right)  $ are defined in terms of
$\left(
A_{i},B_{i}\right)  $ as%

\begin{equation}%
\begin{split}
x_{i}  &  =A_{i}e^{-\alpha_{i}\frac{L}{2}}+B_{i}e^{\alpha_{i}\frac{L}{2}},\\
y_{i}  &  =\alpha_{i}\left(  A_{i}e^{-\alpha_{i}\frac{L}{2}}-B_{i}%
e^{\alpha_{i}\frac{L}{2}}\right)  .
\end{split}
\label{XY}%
\end{equation}
By combining all these conditions (for details see the Appendix), we arrive at
a set of compact formulas for the transmission coefficients
\begin{subequations}
\label{T1T2}%
\begin{align}
T_{1}  &  =\frac{2M_{22}^{\left(  +\right)  }}{M_{11}^{\left(  +\right)
}M_{22}^{\left(  +\right)  }-M_{12}^{\left(  +\right)  }M_{21}^{\left(
+\right)  }}e^{-ik_{1}L},\label{T1}\\
T_{2}  &  =-\frac{2M_{21}^{\left(  +\right)  }}{M_{11}^{\left(  +\right)
}M_{22}^{\left(  +\right)  }-M_{12}^{\left(  +\right)  }M_{21}^{\left(
+\right)  }}e^{-i\left(  k_{1}+k_{2}\right)  L/2}, \label{T2}%
\end{align}
\end{subequations}
and for the reflection coefficients
\begin{subequations}
\label{R1R2}%
\begin{align}
R_{1}  &  =\frac{M_{11}^{\left(  -\right)  }M_{22}^{\left(  +\right)  }%
-M_{12}^{\left(  -\right)  }M_{21}^{\left(  +\right)  }}{M_{11}^{\left(
+\right)  }M_{22}^{\left(  +\right)  }-M_{12}^{\left(  +\right)  }%
M_{21}^{\left(  +\right)  }}e^{-ik_{1}L},\label{R1}\\
R_{2}  &  =\frac{M_{21}^{\left(  -\right)  }M_{22}^{\left(  +\right)  }%
-M_{22}^{\left(  -\right)  }M_{21}^{\left(  +\right)  }}{M_{11}^{\left(
+\right)  }M_{22}^{\left(  +\right)  }-M_{12}^{\left(  +\right)  }%
M_{21}^{\left(  +\right)  }}e^{-i\left(  k_{1}+k_{2}\right)  L/2}, \label{R2}%
\end{align}
\end{subequations}
where
\begin{equation}%
\begin{split}
M_{ij}^{\left(  \pm\right)  }  &  =\sum_{n=1,2}U_{in}\left(  U^{-1}\right)
_{nj} \left[  {\LARGE ~}\left(  1\pm\frac{k_{j}}{k_{i}}\right)  \times~\right.
\\
&  \left.  \cosh\left(  \alpha_{n}L\right)  -i\frac{k_{i}k_{j}\mp\alpha
_{n}^{2}}{k_{i}\alpha_{n}}\sinh\left(  \alpha_{n}L\right)  \right]  .
\end{split}
\label{Mij}%
\end{equation}
In the next section, Eqs. (\ref{T1}) and (\ref{R1}) will be used in connection
with the results from Sec. \ref{model} to numerically determine the lateral shifts.

\section{ \ Discussion}

\label{discussion}

\begin{figure}[h]
\centering
\resizebox{8cm}{!}{\includegraphics{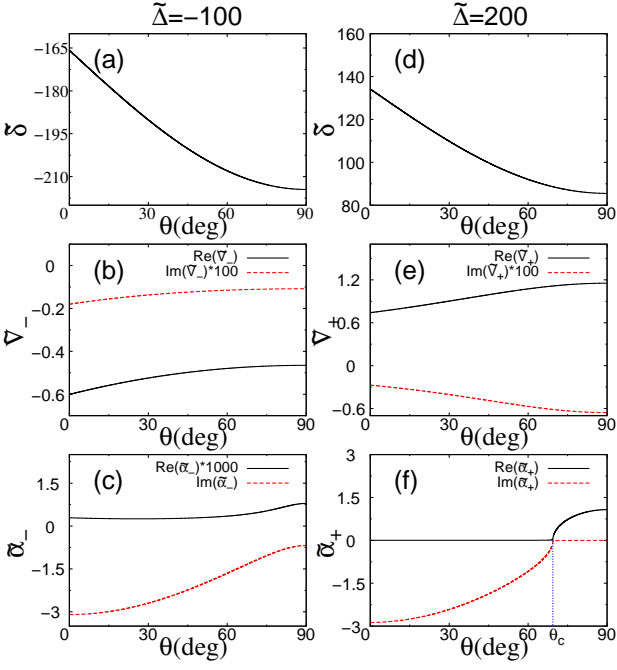}}\caption{(Color
online). The left column calculated under $\tilde{\Delta}=-100$
contains (a) $\tilde{\delta}$, (b) $\tilde{V}_{-}$ , and (c)
$\tilde{\alpha}_{-}$ as functions of $\theta$. \ The right column
calculated under $\tilde{\Delta}=200$ contains (d) $\tilde{\delta}$,
(e) $\tilde{V}_{+}$ , and (f) $\tilde{\alpha}_{+}$ as functions of
$\theta$. \ Other parameters are $\tilde{\gamma}=1,\tilde{k}$
$=3,$ $\tilde{L}=6,~\tilde{\Omega}=20$, and $\tilde{k}_{L}=8.1125$.}%
\label{Fig:V}%
\end{figure}

In this section, we carry out a numerical study of the lateral shifts by first
obtaining $\partial\Theta_{1}^{S}/\partial k_{i}$ from Eq. (\ref{theta_RT})
with the help of Eqs. (\ref{T1}) and (\ref{R1}), and then determining the
lateral shifts using Eqs. (\ref{shiftt}) and (\ref{shifts}). \ In our
calculation, we\ replace $k_{x}=k\cos\theta$ and $k_{y}=k\sin\theta$, and
correspondingly, $\partial/\partial k_{x}=\cos\theta\partial/\partial
k-k^{-1}\sin\theta\partial/\partial\theta$ and $\partial/\partial k_{y}%
=\sin\theta\partial/\partial k+k^{-1}\cos\theta\partial/\partial\theta$, where
$k$ is the magnitude of wavevector and $\theta$ is the incident angle of the
atomic wave. \ In addition, we adopt the following scaled variables:
$\tilde{\Delta}=\Delta/\gamma,\tilde{\Omega}=\Omega/\gamma,$ $\tilde{k}%
_{L}=k_{L}/k_{\gamma}$, $\hat{k}=k/k_{\gamma}$, $\tilde{L}=L/k_{\gamma}^{-1}$,
$\tilde{V}_{\pm}=V_{\pm}/\hbar\gamma$, $\tilde{\alpha}_{\pm}=\alpha_{\pm
}/k_{\gamma}$, and $\tilde{y}_{S}=y_{S}/k_{\gamma}^{-1}$, where $k_{\gamma
}\equiv\sqrt{2m\gamma/\hbar}$. \ In all the examples given bellow, unless
stated otherwise, $\tilde{\gamma}=1,\tilde{k}$ $=3,$ $\tilde{L}=6,~\tilde
{\Omega}=20$, and $\tilde{k}_{L}=8.1125$.

Let's first consider a case where the laser detuning is set at $\tilde{\Delta
}=-100$. \ At this $\tilde{\Delta}$, the effective laser detuning $\delta$
remains (deeply) red detuned across all the incident angles as shown in Fig.
\ref{Fig:V}(a). \ Under such a circumstance, we have $\left\vert
V_{+}\right\vert >>$ $\left\vert V_{-}\right\vert $ according to Eq.
(\ref{Vn}) and $\varphi$ approaches a small value according to Eq.
(\ref{alpha}). \ As a result, the scattering properties in this case are
largely determined by the $\left\vert -\right\rangle $ dressed state. \ For
this reason, we only display in Fig. \ref{Fig:V}(b) potential $V_{-}$ as
\textquotedblleft seen\textquotedblright\ by the atoms in state $\left\vert
-\right\rangle $, which corresponds to a potential well since $Re(V_{-})$
remains negative. \ As a result, the $\left\vert -\right\rangle $ mode
function oscillates in the $x$ dimension with a spatial frequency of
$Im\left(  \alpha_{-}\right)  $, whose value is shown in Fig. \ref{Fig:V}(c)].

\begin{figure}[h]
\centering
\resizebox{8cm}{!}{\includegraphics{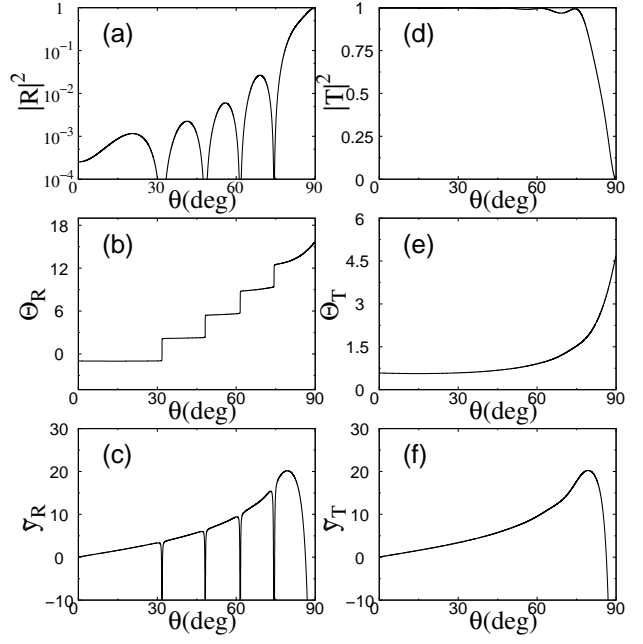}}\caption{The left
column contains (a) $\left\vert R_{1}\right\vert ^{2}$, (b)
$\Theta_{1}^{R}$, and (c) $\tilde{y}_{R}$ as functions of $\theta$.
\ The right column contains (d) $\left\vert T_{1}\right\vert ^{2}$,
(e) $\Theta_{1}^{T}$, and (c) $\tilde{y}_{T}$ as functions of
$\theta$ under the condition that $\tilde{\Delta}=-100$ and the rest
parameters are the same as in Fig.
\ref{Fig:V}.}%
\label{FigN100}%
\end{figure}

The corresponding intensity $\left\vert S_{1}\right\vert ^{2}$, phase
$\Theta_{1}^{S}$, and lateral shift $y_{S}$ of the reflected ($S=R$, left
column) and transmitted ($S=T$, right column) waves are displayed in Fig.
\ref{FigN100} as functions of the incident angle $\theta$. As our discussion
above suggests,\ we expect to see the phenomenon of resonance scattering by
potential wells. \ Indeed, in the region where the phase experiences a $\pi$
shift [Fig. \ref{FigN100}(b)], the reflectivity $\left\vert R\right\vert ^{2}%
$[Fig. \ref{FigN100}(a)] approaches a small value while the transmissivity
$\left\vert T\right\vert ^{2}$ [Fig. \ref{FigN100}(d)] becomes nearly perfect,
which explains the oscillatory behavior exhibited both in $\left\vert
R\right\vert ^{2}$ and in $\left\vert T\right\vert ^{2}$. \ It needs to be
stressed that the phase of the reflected wave, as shown in Fig. \ref{FigN100}%
(b), increases (or decreases) sharply, as $\theta$ (or $k_{x}$) sweeps across
each resonance. \ This causes the reflected wave to experience a large
negative Goos-H\"{a}nchen shift around each resonance as indicated in Fig.
\ref{FigN100}(c).

Next, we consider a situation where $\tilde{\Delta}=200$. \ With this
$\tilde{\Delta}$, $\delta$ remains (deeply) blue detuned for all the incident
angles as Fig. \ref{Fig:V}(d) illustrates. \ As a result, according to Eqs.
(\ref{Vn}) and (\ref{alpha}), $\left\vert V_{+}\right\vert <<\left\vert
V_{-}\right\vert $ and $\varphi$ approaches $\pi/2$; the scattering behavior
is dominated by the $\left\vert +\right\rangle $ dressed state. \ \ Here, the
laser beam, as illustrated in Fig. \ref{Fig:V}(e), creates an effective
repulsive potential $Re\left(  V_{+}\right)  >0$. Under such a circumstance
and provided that $\hbar^{2}k^{2}/2m>Re\left(  V_{+}\right)  $, we can
introduce a critical angle defined as $\theta_{c}\equiv\cos^{-1}%
\sqrt{2mRe\left(  V_{+}\right)  /\hbar^{2}k^{2}}$ at which the ($x$) kinetic
energy of the atomic beam equals the height of the potential barrier. (Such a
critical angle does not exist for the case of red detuning.) $\ $In our case
here, we identify from Fig. \ref{Fig:V}(f) that $\theta_{c}\approx69.4^{\circ
}$, which has the physical meaning that bellow $\theta_{c}$, the $\left\vert
+\right\rangle $ mode oscillates at a spatial frequency close to $Im\left(
\alpha_{+}\right)  $ while beyond $\theta_{c}$, it undergoes quantum tunneling
with $1/Re\left(  \alpha_{+}\right)  $ being the characteristic tunneling distance.

\begin{figure}[h]
\centering
\resizebox{8cm}{!}{\includegraphics{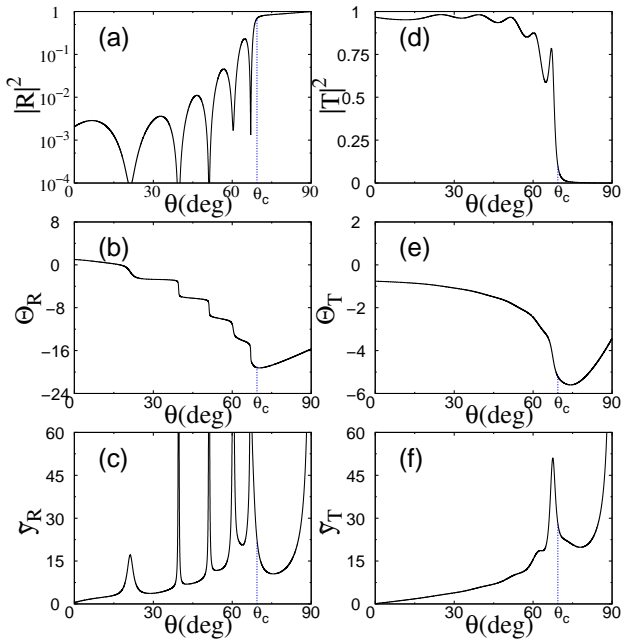}}\caption{(Color
online). The left column contains (a) $\left\vert R_{1}\right\vert
^{2}$, (b) $\Theta_{1}^{R}$, and (c) $\tilde{y}_{R}$ as functions of
$\theta$. \ The right column contains (d) $\left\vert
T_{1}\right\vert ^{2}$, (e) $\Theta_{1}^{T}$, and (c)
$\tilde{y}_{T}$ as functions of $\theta$ under the condition that
$\tilde{\Delta}=200$ and the rest parameters are the same as in Fig.
\ref{Fig:V}.}%
\label{Fig:P200}%
\end{figure}

Indeed, Fig. \ref{Fig:P200} shows that $\left\vert S_{1}\right\vert ^{2}$,
$\Theta_{1}^{S}$, and $y_{S}$ beyond the critical angle $\theta>\theta_{c}$
are qualitatively different from those within the critical angle
$0<\theta<\theta_{c}$. For $\theta>\theta_{c}$, besides relatively sharp
features around the boundaries, both the phase and lateral shift exhibit no
oscillations. \ This description applies both to the reflected and transmitted
beams. \ \ For $0<\theta<\theta_{c}$ where the incident ($x$) kinetic energy
exceeds the potential height, the reflection [Fig. \ref{Fig:P200}(a)] and
transmission [Fig. \ref{Fig:P200}(d)] oscillate. \ The phase of the reflected
wave, while still undergoes a $\pi$ shift, decreases (or increases) sharply,
as $\theta$ (or $k_{x}$) sweeps across each resonance [Fig. \ref{Fig:P200}%
(b)], a behavior completely opposite to the case of red detuning [Fig.
\ref{FigN100}(b)]. \ As a result, we see that the reflected wave develops a
large but positive Goos-H\"{a}nchen-like lateral shift around each resonance
[Fig. \ref{Fig:P200}(c)] \ (This is in contrast to the phase and lateral shift
of the transmitted wave, which remain relatively monotonic within the critical angle.)

It needs to be emphasized that large negative shifts around resonance have
been the focus of several recent studies for light waves that propagate
through absorptive medium slabs in conventional optics
\cite{WangLG:1,Lai:1,Wild}. \ \ In our atom optics model here, the
ground-state matter wave is coupled to the excited-state matter wave, and this
coupling greatly enriches the physics concerning the lateral shifts. \ Not
only do we see large negative lateral shifts as in Fig. \ref{FigN100}(c) when
the laser is red detuned, but also large positive lateral shifts as in Fig.
\ref{Fig:P200}(c) when the laser is blue detuned. \ Moreover, with our atom
optics model, great controls over not only the position but also the linewidth
of these peak shifts can be achieved by taking advantage of lasers being
highly tunable both in intensity and in frequency (not shown).

\section{Conclusion}

\label{conclusion}

In conclusion, we have established a theoretical framework for studying the
matter wave analog of Goos-H\"{a}nchen-like effect in an atom optics model
where a super-Gaussian laser beam acts as a "medium slab" for a matter wave of
two-level atoms. \ We have developed \ a vector theory based upon a set of
coupled Schr\"{o}dinger equations for describing the scattering of a
wavepacket of two-level atoms off a square potential. We have derived a set of
analytical formulas for the transmission and reflection coefficients, which
have greatly facilitated the study of Goos-H\"{a}nchen effect in vector models
where atoms are treated as particles possessing two internal spin components.
\ \ It is important to stress that the coupling between the ground and excited
components in the vector model combined with the tunability offered by the
laser field creates new opportunities for studying the lateral shifts. \ In
particular, we have found that in our atom optics model,\ not only a large
negative Goos-H\"{a}nchen shift as in conventional optics
\cite{WangLG:3,Lai:1,Shen,Yan} but also a large positive shift can take place
in the reflected atomic beam. \

\section{Acknowledgements}

\textbf{We thank H. Pu for helpful discussion.} This work is
supported by the National Natural Science Foundation of China under
Grant No. 10588402 and No. 10474055, the National Basic Research
Program of China (973 Program) under Grant No. 2006CB921104, the
Science and Technology Commission of Shanghai Municipality under
Grant No. 06JC14026, No. 05PJ14038, the Program for Changjiang
Scholars and Innovative Research Team in University, Shanghai
Leading Academic Discipline Project No. B408, the Research Fund for
the Doctoral Program of Higher Education No. 20040003101, and by the
US National Science Foundation
and the US Army Research Office (HYL).%

\begin{appendix}%

\section{Appendix}

In this Appendix, we provide the steps leading to Eqs. (\ref{T1T2}) and
(\ref{R1R2}). \ To begin with, we insert, $A_{i}=\frac{1}{2}\left(
u_{i}+\frac{v_{i}}{\alpha_{i}}\right)  e^{-\alpha_{i}L/2}$and $B_{i}=\frac
{1}{2}\left(  u_{i}-\frac{v_{i}}{\alpha_{i}}\right)  e^{\alpha_{i}L/2}$
obtained from Eqs. (\ref{ui vi}), into Eqs. (\ref{XY}), enabling us to express
$\left(  x_{i}\text{,}y_{i}\right)  $ in terms of ($u_{i}$, $v_{i}$) as%

\begin{subequations}
\label{XY2}%
\begin{align}
x_{i}  &  =u_{i}\cosh\left(  \alpha_{i}L\right)  -v_{i}\sinh\left(  \alpha
_{i}L\right)  /\alpha_{i},\\
y_{i}  &  =v_{i}\cosh\left(  \alpha_{i}L\right)  -\alpha_{i}u_{i}\sinh\left(
\alpha_{i}L\right)  .
\end{align}
\end{subequations}
By combining Eq. (\ref{X}) and Eq. (\ref{Y}), we eliminate $R_{1}$
and $R_{2}$ simultaneously from Eqs. (\ref{XY1}) and arrive at a
single-matrix equation
\begin{equation}
U\left(
\begin{array}
[c]{c}%
x_{1}\\
x_{2}%
\end{array}
\right)  +W\left(
\begin{array}
[c]{c}%
y_{1}\\
y_{2}%
\end{array}
\right)  =\left(
\begin{array}
[c]{c}%
2e^{-ik_{1}L/2}\\
0
\end{array}
\right)  ,
\end{equation}
which, by virtue of Eqs. (\ref{uv}) and (\ref{XY2}), is shown to be equivalent
to
\begin{equation}
M^{\left(  +\right)  }\left(
\begin{array}
[c]{c}%
T_{1}e^{ik_{1}L/2}\\
T_{2}e^{ik_{2}L/2}%
\end{array}
\right)  =\left(
\begin{array}
[c]{c}%
2e^{-ik_{1}L/2}\\
0
\end{array}
\right)  , \label{T equations}%
\end{equation}
where $M^{\left(  +\right)  }$ is a $2\times2$ matrix given by%
\begin{widetext}%
\begin{align}
M^{\left(  \pm\right)  }  &  =U\left[
\begin{array}
[c]{cc}%
\cosh\left(  \alpha_{1}L\right)  & 0\\
0 & \cosh\left(  \alpha_{2}L\right)
\end{array}
\right]  U^{-1}-U\left[
\begin{array}
[c]{cc}%
\sinh\left(  \alpha_{1}L\right)  /\alpha_{1} & 0\\
0 & \sinh\left(  \alpha_{2}L\right)  /\alpha_{2}%
\end{array}
\right]  W^{-1}\nonumber\\
&  \pm W\left[
\begin{array}
[c]{cc}%
\cosh\left(  \alpha_{1}L\right)  & 0\\
0 & \cosh\left(  \alpha_{2}L\right)
\end{array}
\right]  W^{-1}\mp W\left[
\begin{array}
[c]{cc}%
\alpha_{1}\sinh\left(  \alpha_{1}L\right)  / & 0\\
0 & \alpha_{2}\sinh\left(  \alpha_{2}L\right)
\end{array}
\right]  U^{-1}. \label{M}%
\end{align}
\end{widetext}
By a similar procedure, we find from Eq. (\ref{XY1}) that
\begin{equation}
2\left(
\begin{array}
[c]{c}%
R_{1}e^{ik_{1}L/2}\\
R_{2}e^{ik_{2}L/2}%
\end{array}
\right)  =U\left(
\begin{array}
[c]{c}%
x_{1}\\
x_{2}%
\end{array}
\right)  -W\left(
\begin{array}
[c]{c}%
y_{1}\\
y_{2}%
\end{array}
\right)  ,
\end{equation}
which, with the help of Eqs. (\ref{uv}) and (\ref{XY2}), is equivalent to
\begin{equation}
\left(
\begin{array}
[c]{c}%
2R_{1}e^{ik_{1}L/2}\\
2R_{2}e^{ik_{2}L/2}%
\end{array}
\right)  =M^{\left(  -\right)  }\left(
\begin{array}
[c]{c}%
T_{1}e^{ik_{1}L/2}\\
T_{2}e^{ik_{2}L/2}%
\end{array}
\right)  , \label{R equations}%
\end{equation}
where $M^{\left(  -\right)  }$ is also $2\times2$ matrix given by Eq.
(\ref{M}). \ A straightforward calculation involving the use of Eq.
(\ref{Wij}) shows that the matrix element $M_{ij}^{\left(  \pm\right)  }$ of
Eq. (\ref{M}) has a simple and explicit form given by Eq. (\ref{Mij}).
Finally, we arrive at Eqs.(\ref{T1T2}) and (\ref{R1R2}) by solving Eqs.
(\ref{T equations}) and (\ref{R equations}) simultaneously.%

\end{appendix}%
%

\setcounter{secnumdepth}{-1}%

*wpzhang@phy.ecnu.edu.cn

\dag\ ling@rowan.edu

\end{document}